\documentclass[pre,twocolumn]{revtex4}
\usepackage{amsmath}
\usepackage{amssymb}
\usepackage{graphicx}

\begin{document}

\title{Unfolding and unzipping of single-stranded DNA by stretching}
\author{Alexei V. Tkachenko (alexei@umich.edu)}
\affiliation{Michigan Center for Theoretical Physics and Department of Physics,
University of Michigan, 500 E. University Ave., Ann Arbor, 48109 MI, USA}

\begin{abstract}
We present a theoretical study of single-stranded DNA under stretching.
Within the proposed framework, the effects of basepairing on the mechanical
response of the molecule can be studied in combination with an arbitrary
underlying model of chain elasticity. In a generic case, \ we show that the
stretching curve \ of ssDNA exhibits two distinct features: the second-order
"unfolding" phase transition, and a sharp crossover, reminiscent of the
first-order "unzipping" transition in dsDNA. We apply the theory to the
particular cases of Worm-like Chain (WLC) and Freely-Joint Chain (FJC)
models, and discuss the universal and model--dependent features of the
mechanical response of ssDNA. In particular, we show that variation of the
width of the unzipping crossover with interaction strength is very sensitive
to the energetics of hairpin loops. This opens a new way of testing the
elastic properties of ssDNA.

\end{abstract}

\maketitle

\section{Introduction\protect\bigskip}

A dramatic progress has been made over last decade in employing the
single-molecule micromanipulation techniques for studies of biological
materials and processes. Pioneered by the work of Smith et. al. \cite%
{stretch} on stretching of double--stranded DNA (dsDNA), these techniques
have later been applied to study proteins, DNA--protein interactions,
chromosomes etc. The chain--stretching experiments were also performed on
single--stranded DNA molecules (ssDNA)\cite{ssDNA1}-\cite{ssDNAexp2}.

One could expect the response of ssDNA to stretching to be dramatically
different from that of dsDNA, because of the effects of the possible
basepairing between complementary segments of the chain. The need for
understanding of the resulting mechanical behavior has already attracted a
considerable attention of theorists \cite{Hwa}-\cite{zhou}. In particular,
it has been shown \cite{Mezard}-\cite{muller} that in the thermodynamic
limit, ssDNA chain should undergo the second--order phase transition, at
finite critical force. Formally, this phenomenon is very similar to the
hypothetical native--molten transition in RNA \cite{hwa-RNA}, as well as to
the classical (yet unconfirmed) picture of dsDNA denaturation \cite{poland}.

The predicted critical behavior is qualitatively different from another
related phenomenon, the force-induced denaturation (unzipping) of dsDNA \cite%
{unzip1}-\cite{unzip}. The unzipping is a first--order transition, which
occurs as a result of competition between the elastic energy of the
stretched ssDNA, and the basepairing (hybridization) energy within dsDNA.
The very same effects are essential for the above second-order phase
transition in stretched ssDNA. As a part of the present work, we will
clarify the relationship between the two phenomena.

The theoretical modelling of ssDNA and RNA is traditionally done within
Freely Joint Chain (FJC) model. Its extensible version has been originally
used for fitting of the early ssDNA stretching data \cite{ssDNA1}. However,
both the microscopic structure of ssDNA, and the recent experiments with DNA
hairpin constructs \cite{libs}-\cite{kuzn} , strongly suggest the picture
with a finite bending modulus of its backbone, which feature is reminiscent
of the semiflexible Worm--Like Chain (WLC) model \cite{semiflex}.
Nevertheless, being a continuous model, the WLC description is unlikely to
be valid in the regimes when discrete nature of chemical bonds becomes
relevant (e.g. for a sufficiently high stretching force). To overcome this
limitation, Discreet Persistent Chain (DPC) model has been proposed for
ssDNA in recent work by Storm and Nelson \cite{pnels}. Interestingly, WLC
itself was originally introduced by Kratky and Porod \cite{semiflex} as a
continuous limit of a similar discreetized model. The authors of \cite{pnels}
have shown, that the extensible versions of all three models (FJC, WLC, and
DPC) produce a good fit to the experimentally observed stretching curves, as
long as force is not too high ($f\lesssim 200pN$ for FJC, and $f\lesssim
400pN$ for WLC/DPC). \ 

The central idea of our paper is to include the effects of basepairing
within a theoretical framework compatible with an arbitrary underlying model
of ssDNA elasticity. In this way we can separate the two parts of the
problem: the search of an adequate elastic description of ssDNA, and the
evaluation of the effects of its self-interactions. In fact, one may be able
to start with an empirical elastic models extracted from independent
experiments, and use our theory to predict the stretching behavior of the
chain after "switching on" the basepairing interactions.

\section{Theoretical Framework}

We consider ssDNA chain subjected to external pulling force $f$. As an input
for our theory, one has to specify two functions characterizing the system
without self--interactions: $q_{el}(f)$ -- elastic free energy per unit
chain length vs. pulling force; and $F_{loop}(l)$ -- free energy of a loop
as a function of its contour length. Our goal is to study the effect of
interactions between complementary segments of the chain. Similarly to ref 
\cite{Mezard}, we assume the interaction strength to be the same \ for any
two chain fragments. It will be characterized by a single parameter, $%
\epsilon $,\ defined as pairing energy per unit chain length. Strictly
speaking, this should limit applicability of our approach to
self--complementary periodic sequences of ssDNA (such as ATATAT...).
However, as discussed in \cite{Mezard}-\cite{Mezard2}, this uniform model
may be reasonably adequate for random sequences, too. Nevertheless, the
effect of randomness remains an important problem for the future studies.

Hybridization of distant chain segments results in looping. Therefore, the
interactions reduce the effective (free) chain length $l$ exposed to the
stretching force. Our goal is to calculate the partition function of the
system $Z(L,l)$, parameterized by the total chain length $L$, and free
length $l$. The general form of the interaction--free partition function is, 
$Z_{0}(L,l)=\exp \left( \mu _{0}L\right) \delta \left( L-l\right) $, because 
$l$ and $L$ coincide, and different chain segments are statistically
independent (we neglect excluded volume effects). Without loss of
generality, one can choose $\mu _{0}=0$. \ It is useful to perform double
Laplace transform of partition function $Z(L,l)$. This results in
introduction of parameters $\mu $ and $q$, conjugated to $L$ and $l$
respectively (parameter \thinspace $\mu $ conjugated to the total length, is
often called fugacity): 
\begin{equation}
Z(\mu ,q)=\int\nolimits_{0}^{\infty }\int\nolimits_{0}^{\infty }Z(L,l)\exp 
\left[ -\mu L-ql\right] \mathrm{d}L\mathrm{d}l.
\end{equation}%
In particular, $Z_{0}(\mu ,q)=1/\left( \mu +q\right) $. Since the elastic
part of free energy is $q_{el}(f)l$, the overall free energy may be
expressed as 
\begin{equation}
F(L,f)=-\log \int\limits_{-i\infty }^{+i\infty }Z\left( \mu
,q_{el}(f)\right) \mathrm{e}^{\mu L}\frac{\mathrm{d}\mu }{2\pi i}.
\end{equation}%
Here and below, we take $k_{B}T=1$.

\begin{figure}[tbp]
\begin{center}
\includegraphics[
height=2.1248in,
width=2.975in
]{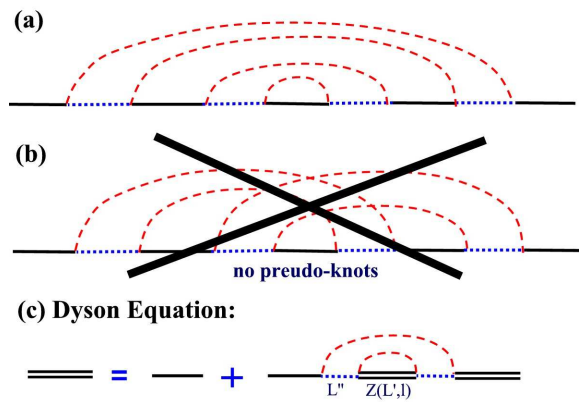}
\end{center}
\caption{Schematic representation of partition function calculation. Thin
solid lines represent $Z_{0}$, while dashed line indicate base-pairing of
distant chain segments (dotted). }
\label{diag}
\end{figure}

It is well known that the partition function of a uniform RNA or ssDNA may
be calculated in a recursive manner, reminiscent of Hartree approximation 
\cite{Hwa}-\cite{hwa-RNA}. This calculation can be represented in a
diagrammatic form shown in Figure \ref{diag}. The solid lines correspond to
bare partition function $Z_{0}$; dotted fragments connected with dashed
lines represent pairing (hybridization) of the corresponding chain segments.
Within our approach, the energy cost of the pairing is $\epsilon
L_{hyb}+F_{loop}(l)$, where $L_{hyb}$ is the length of the hybridized
region, and $l$ is the free length of the chain segment being "internalized"
due to the looping. In fact, one can extend our approach to a more general
form of hybridization energy, $\varepsilon _{0}+\epsilon L_{hyb}$, with
constant $\varepsilon _{0}>0$ representing the energy cost of termination of
the double--stranded segment. Introduction of $\varepsilon _{0}$ accounts
for the cooperativity of the basepairing interactions.

Traditionally, problems involving DNA/RNA folding or denaturation, are
studied within a discreet model. Each discrete "monomer" in this approach
represents a chain segment which can hybridize independently of its
neighbors. Its length is assumed to be equal to the statistical segment,
which makes it easy to combine this description with FJC model. Since we are
interested in \ developing a theory for an arbitrary model of chain
elasticity, it is logical to abandon this artificial discretization of
hybridized segments. Of course, the basepairing remains fundamentally
discreet on the length scale $l_{0}$ of a single base. As long as the
relevant physics occurs on larger scales, ssDNA can be considered as a
continuous chain.

The crucial observation is that one can typically neglect all the diagrams
with intersecting dashed lines. Such situation would correspond to
"pseudo--knot", whose probability is low because it requires winding of
ssDNA/RNA around itself (Fig. \ref{diag}b). Thanks to this topological
constrain, "self energy" diagram entering Dyson equation (Fig \ref{diag}c),
may be calculated exactly within one-loop (Hartree) approximation. Since we
have assumed the uniform interaction parameter, the problem has a
particularly simple form at the Laplace representation: 
\begin{align}
Z^{-1}(\mu ,q)& =Z_{0}^{-1}(\mu ,q)-\iiint \frac{\mathrm{d}L^{\prime }%
\mathrm{d}L^{\prime \prime }\mathrm{d}l}{l_{0}^{3}}Z\left( L^{\prime
},l\right) \times  \label{dyson} \\
& \times \exp \left[ -\mu \left( 2L^{\prime \prime }+L^{\prime }\right)
-\epsilon L^{\prime \prime }-F_{loop}(l)-\varepsilon _{0}\right] .  \notag
\end{align}%
From here, one can obtain: 
\begin{equation}
Z^{-1}(\mu ,q)=\mu +q-\frac{W\left( q_{\mu }\right) }{2\mu +\epsilon },
\end{equation}%
were $q_{\mu }$ corresponds to the pole in $Z(\mu ,q)$, and 
\begin{equation}
W\left( q\right) =\frac{\mathrm{e}^{-\varepsilon _{0}}}{l_{0}^{3}}%
\int\limits_{0}^{\infty }\exp \left[ -F_{loop}(l)+ql\right] \mathrm{d}l.
\label{Wq}
\end{equation}

\section{Generic behavior: "unzipping" versus "unfolding"\protect\bigskip}

In the thermodynamic limit ($L\longrightarrow \infty $),%
\begin{equation}
l=-\frac{\partial }{\partial q}\left. \log \int\limits_{-i\infty }^{+i\infty
}Z\left( \mu ,q\right) \mathrm{e}^{\mu L}\frac{\mathrm{d}\mu }{2\pi i}%
\right\vert _{q_{el}\left( f\right) }=-L\left. \frac{\mathrm{d}\mu _{q}}{%
\mathrm{d}q}\right\vert _{q_{el}\left( f\right) }.  \label{lf}
\end{equation}%
Here function $\mu _{q}\left( q\right) $ again corresponds to the pole in $%
Z(\mu ,q)$, i.e. it is inverse to $q_{\mu }\left( \mu \right) $. Explicitly,%
\begin{equation}
\mu _{q}\left( q\right) =-\frac{1}{2}\left[ \frac{\epsilon }{2}+q-\sqrt{%
\left( \frac{\epsilon }{2}-q\right) ^{2}+2W\left( q\right) }\right] .
\label{mu}
\end{equation}%
This yields the following result for the free length as a function of
tension:%
\begin{equation}
\frac{l}{L}=\frac{1}{2}\left( 1-\frac{q_{el}-\epsilon /2+W^{\prime }\left(
q_{el}\right) }{\sqrt{\left( q_{el}-\epsilon /2\right) ^{2}+2W\left(
q_{el}\right) }}\right) .  \label{lf2}
\end{equation}%
Here $W^{\prime }\equiv \mathrm{d}W/\mathrm{d}q$. Note that from the
definition of function $W$, Eq. (\ref{Wq}), one can relate its logarithmic
derivative to the average length of the loop:%
\begin{equation}
l_{loop}(q)=\frac{W^{\prime }}{W}
\end{equation}

If $l$ is known, one can easily find the relative elongation of the chain,
i.e. the ratio of the end-to-end distance $R$ to the total chain length $L$,
which is observable experimentally:%
\begin{equation}
\frac{R}{L}=xl=-\frac{1}{2}\frac{\partial q_{el}}{\partial f}\left( 1-\frac{%
q_{el}-\epsilon /2+W^{\prime }\left( q_{el}\right) }{\sqrt{\left(
q_{el}-\epsilon /2\right) ^{2}+2W\left( q_{el}\right) }}\right)  \label{RL}
\end{equation}%
Here $x=-\partial q_{el}/\partial f$ is the relative elongation of the free
portion of the chain.

According to Eq. (\ref{lf2}), in the strong stretching limit ($%
q_{el}\rightarrow -\infty $) the chain becomes completely "free": $%
l\rightarrow L$. Further examination of our expression for the free length
reveals its peculiar behavior near the point $q_{el}=\epsilon /2$. In fact,
in the limit of vanishing $W$ (i.e. very high energy cost of a loop), $%
l\left( q_{el}\right) $ becomes a step function changing from $0$ to $L$ at
that point. In a realistic situation, it is transformed into a crossover
whose width depends on $W$, as%
\begin{equation}
\delta _{q}=\sqrt{2W\left( \epsilon /2\right) }.
\end{equation}%
Since the location of that crossover corresponds to the point were the
hybridization free energy is exactly equal to the elastic one, we conclude
that this behavior is directly related to the first-order \textit{unzipping}
transition \ of dsDNA. The transition is transformed into the crossover due
to the finite size of the hybridized segments.

Another \ important feature of our result is that the free fraction of the
chain $l/L$ goes to zero at finite tension $q_{el}=q^{\ast }$. According to
Eq. (\ref{lf2}) the condition for this to happen is, 
\begin{eqnarray}
q^{\ast }-\epsilon /2 &=&\frac{W\left( q^{\ast }\right) }{W^{\prime }\left(
q^{\ast }\right) }-\frac{W^{\prime }\left( q^{\ast }\right) }{2}=
\label{transition} \\
&&\frac{1}{l_{loop}(q^{\ast })}-\frac{l_{loop}(q^{\ast })}{l_{0}^{2}}W\left(
q^{\ast }\right)  \notag
\end{eqnarray}%
This point corresponds to the second order phase transition, which has been
reported in the earlier studies of the problem. The crucial observation is
that this transition is physically distinct from the unzipping crossover. It
may be viewed as a precursor of the force-induced denaturation. Indeed, even
when the tension in the free segments is still insufficient to overcome the
binding energy $-\epsilon $, one can convert a finite fraction of the chain
into the free length without breaking any bonds. Namely, since there is a
finite density of loops at the folded ssDNA, one can simply "move" the
unbound bases from the loops to the free segment. \ This will only result in
an entropy loss, but will not reduce the interaction energy.

Indeed, if the basepairing energy is negative, the magnitude of the critical
tension is somewhat lower ($\left\vert q^{\ast }\right\vert <-\epsilon /2$ )
than that at the unzipping point: 
\begin{equation}
q^{\ast }-\frac{\epsilon }{2}\approx \frac{1}{l_{loop}(q^{\ast })}
\label{qstar}
\end{equation}

The free segments of the chain exposed to the external stretching can be
viewed as topologically distinct objects ("vortices"), which separates the
system onto independently folded domains. The effective interaction
potential of two consequent vortices is determined by the free energy of the
folded region between them. \ At the point $q^{\ast }$, the typical
separation between the vortices along the chain coordinate, i.e. the size of
the separable domains, diverges. This \textit{folding-unfolding} transition
occurs as a result of competition between the elastic tension and folding
entropy, which grows with the domain size.

Interestingly, Eq. (\ref{transition}) has physical solutions even for
positive interaction energy, i.e. when the basepairing is energetically
unfavorable. It follows form that fact that for $l_{loop}\longrightarrow
\infty $, the generic behavior of looping free energy is $F_{loop}(l)\simeq
3/2\log l$, (for an ideal Gaussian chain). As a result, $W\left( q\right) $
remains finite for $q\longrightarrow 0$ , while $W^{\prime }$ and $%
l_{loop}(q^{\ast })$ are diverging, as $q^{\nu -2}$. Therefore, according to
Eq. (\ref{transition}), there is a finite tension at which the unfolding
transition takes place, for $\epsilon >0$. The asymptotic relationship
between $\epsilon $ and $q^{\ast }$ in this regime is,

\begin{equation}
l_{loop}(q^{\ast })\approx \frac{\epsilon l_{0}^{2}}{2}W\left( 0\right) 
\end{equation}%
This yields a power law dependence of the critical tension on $\epsilon $: $%
q^{\ast }\sim \epsilon ^{-2}$. Note however that our conclusion about the
unfolding transition at positive values of basepairing energy, may well be a
result of limitations of the model.

\section{Application to Worm-Like Chain model}

As an example, we discuss the application of the above approach to the
particular case of WLC model. Its Hamiltonian is given by 
\begin{equation}
H=\int\limits_{0}^{L}\left[ \frac{l_{p}}{2}\left( \frac{\partial ^{2}\mathbf{%
r}}{\partial s^{2}}\right) ^{2}\right] \mathrm{d}s
\end{equation}%
were $l_{p}$ is persistence length, and $\mathbf{r}\left( s\right) $ defines
the spatial conformation of the chain, subjected to constrain $\left\vert
\partial \mathbf{r/}\partial s\right\vert =1$. For this model, Marko and
Siggia \cite{semiflex} have proposed the following interpolative
relationship between the stretching force $f$ and relative extension $%
x\equiv R/L$. 
\begin{equation}
f\left( x\right) =\frac{1}{l_{p}}\left( \frac{1}{4}\left[ \frac{1}{\left(
1-x\right) ^{2}}-1\right] +x\right)
\end{equation}%
From here, function $q_{el}(f)$ (which is one of the inputs for our theory),
can be expressed in parametric form: 
\begin{equation}
q_{el}(f)=\int\limits_{0}^{x}f(x^{\prime })dx^{\prime }-xf\left( x\right) =-%
\frac{x^{2}}{4l_{p}}\left[ \frac{1}{\left( 1-x\right) ^{2}}+2\right]
\end{equation}

\begin{figure}[tbp]
\begin{center}
\includegraphics[
height=2.5192in,
width=3.3183in 
]{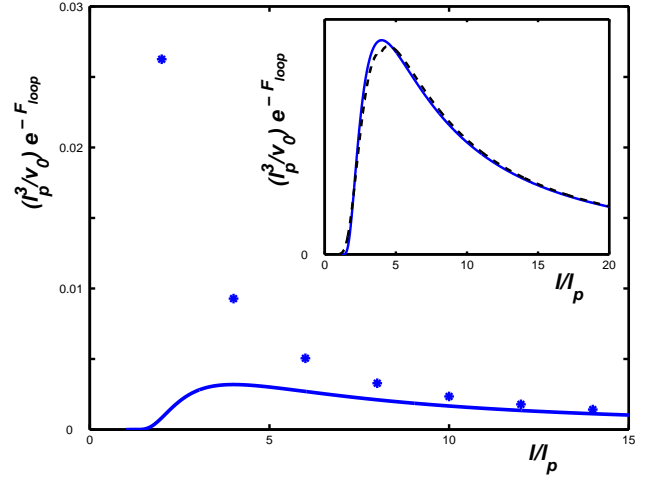}
\end{center}
\caption{Looping probability, $\exp \left( -F_{loop}\right) $, as a function
of loop size for different models of ssDNA elasticity. Here the solid line
represents the analytic result for WLC, Eq. (\protect\ref{loopsimple}), and
points correspond to FJC model. Insert: comparison of classical
interpolative result for looping probability (from Ref. \protect\cite%
{Yamakawa}, dashed line), with the simplified formula, Eq. (\protect\ref%
{loopsimple}).}
\label{F_loop}
\end{figure}

One has also to specify the loop free energy. \ An analytic interpolation
for $F_{loop}$ was proposed in Ref. \cite{Yamakawa}, in the context of the
ring cyclization problem: 
\begin{equation}
\mathrm{e}^{-F_{loop}}\simeq \frac{v_{0}}{l_{p}^{3}}\left\{ 
\begin{array}{c}
4\pi ^{3}\left( \frac{2l_{p}}{l}\right) ^{6}\exp \left( -\frac{2\pi ^{2}l_{p}%
}{l}+\frac{l}{4l_{p}}\right) ,\text{ }l<4l_{p} \\ 
\left( \frac{3l_{p}}{\pi l}\right) ^{3/2}\left[ 1-\frac{11}{4}\frac{l_{p}}{l}%
+\frac{l_{p}^{2}}{5l^{2}}\right] ,\text{ \ }l>4l_{p}%
\end{array}%
\right.   \label{Floop}
\end{equation}%
Here $v_{0}\lesssim 1\mathring{A}^{3}$ is the effective "reaction volume"
associated with the localization of the loop ends by the hydrogen bonding.
We have found a simpler version of the interpolative expression, which gives
a very good fit to the global behavior of the looping probability, $\mathrm{e%
}^{-F_{loop}}$ (see Figure \ref{F_loop}): 
\begin{equation}
\mathrm{e}^{-F_{loop}}\simeq v_{0}\left( \frac{3}{\pi l_{p}\left(
l-l_{p}\right) }\right) ^{3/2}\exp \left( -\frac{4.5}{l/l_{p}-1}\right)
\theta \left( l-l_{p}\right)   \label{loopsimple}
\end{equation}%
Here $\theta $ is the step function.

Within this approximation, function $W_{WLC}\left( q\right) $ can be found
analytically: 
\begin{equation}
W\left( q\right) \simeq \frac{\alpha }{l_{p}^{2}}\exp \left[ -3\sqrt{-2ql_{p}%
}+ql_{p}\right]
\end{equation}%
Here 
\begin{equation}
\alpha =\frac{\sqrt{6}}{\pi }\frac{v_{0}}{l_{0}^{3}}\mathrm{e}^{-\varepsilon
_{0}}
\end{equation}%
Note that $\alpha $ replaces three parameters of the original model, $v_{0} $%
, $l_{0}$, and energy $\varepsilon _{0}$. Since $\delta \lesssim l_{0}$, and 
$\varepsilon _{0}>0$, this dimensionless parameter is expected to be small.
It has a physical meaning of the effective reaction volume of the loop ends,
in units of $l_{0}^{3}$.

Now that we have specified $W_{WLC}\left( q\right) $ and $q(f)$, Eqs. (\ref%
{lf})-(\ref{mu}), can be used to find $l(f)$, together with force--extension
curves, $R(f)=l(f)x(f)$. The fact that $\alpha $ is a small parameter allows
us to simplify the results. In the regime of negative $\epsilon $, a sharp
unzipping crossover is expected. As we have discussed in the previous
section, its characteristic width is given by 
\begin{equation}
\delta _{q}=\sqrt{2W\left( \epsilon /2\right) }=\frac{\sqrt{2\alpha }}{l_{p}}%
\exp \left[ \left( -6\sqrt{-\epsilon l_{p}}+\epsilon l_{p}\right) /4\right]
\label{deltaq}
\end{equation}%
In the vicinity of the crossover point, $q_{el}=\epsilon /2$, the change in
free length can be well described by the universal function of rescaled
variable $\Delta =\left( q-\epsilon /2\right) /\delta _{q}$: 
\begin{equation}
\frac{l}{L}\simeq \frac{1}{2}\left( 1-\frac{\Delta }{\sqrt{\Delta ^{2}+1}}%
\right)  \label{lf-universal}
\end{equation}%
For large enough $\Delta $, this universal behavior breaks down due to the
proximity of the unfolding phase transition. The corresponding critical
tension $q^{\ast }$ is given by:%
\begin{equation}
q_{WLC}^{\ast }\simeq \frac{\epsilon }{2}+\frac{1}{l_{loop}(\epsilon /2)}%
\simeq \frac{\epsilon }{2}+\frac{1}{l_{p}\left( 1+3/\sqrt{-\epsilon l_{p}}%
\right) }  \label{qstarwlc}
\end{equation}

\begin{figure}[tbp]
\begin{center}
\includegraphics[
height=2.6496in,
width=3.3036in
]{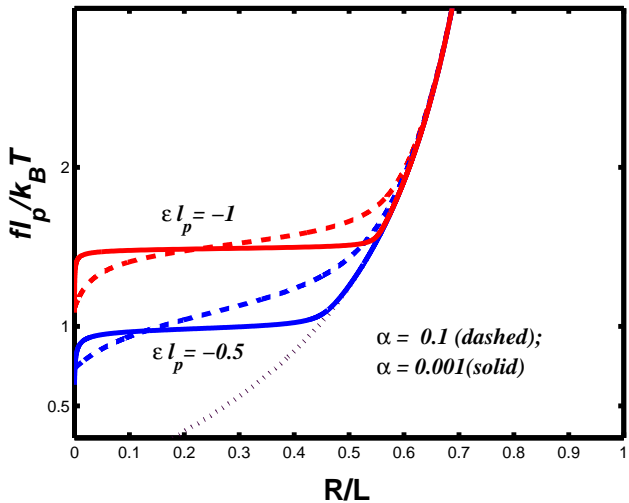}
\end{center}
\caption{Theoretical stretching curves for WLC model with self-interaction.
Dotted line corresponds to interaction-free WLC.}
\label{pulling}
\end{figure}

Figure \ref{pulling} shows the sharp \emph{\ \ }crossover at $\left\vert
q(f)\right\vert =\epsilon /2$. Consistently with Eq. (\ref{deltaq}), its
sharpness increases with lowering parameter $\alpha $. This can be
associated with growth of the typical length of hybridized regions. The
unzipping is clearly separated from second--order unfolding transition.
While this crossover regime was not present at the early studies of the
problem, it has been recently reported in Ref. \cite{muller}. In that work,
the traditional FJC-based model has been modified to include the effects of
cooperativity, analogous to our parameter $\varepsilon _{0}$. As we have
discussed, the sharp crossover should be interpreted as unzipping (force
induced denaturation), while the second-order transition corresponds to
topological change (unfolding) which may be viewed as a precursor of the
unzipping. This physical picture is consistent with the results of Ref. \cite%
{muller}. On the other hand, our analysis disagrees with the conclusions of
Ref. \cite{zhou}, in which the first order phase transition was predicted
for the regime of strong enough hybridization energy.

\section{Model dependence of the results}

Here we discuss how the above results may depend on the choice of the
elastic description of ssDNA. It should be noted that such important
features, as the unfolding transition and the unzipping crossover, are very
robust and nearly independent of the model. Furthermore, it is well known
the stretching curve of ssDNA may be fitted reasonably well by several
models, e.g. \ extensible versions of FJC, WLC, or DPC. This implies that
deduction of $q_{el}\left( f\right) $ from the existing and future
experimental data is unlikely to provide a sensitive test for the possible
models. On the other hand, we have seen that the loop free energy $%
F_{loop}\left( l\right) $ and consequently $W\left( q\right) $ are
significant parameters of the problem. As we shall see, this parameters are
very sensitive to the choice of the underlying elastic model.

In the case of Discreet Persistent Chain (DPC) model, we do not expect any
significant deviation of $F_{loop}\left( l\right) $ or $W\left( q\right) $
from those obtained for WLC model, since the typical bending radius
significantly exceeds the bond length. \ The (extensible) FJC model has been
a standard framework in which the discussed problem has been studied so far.
Nevertheless, here we briefly review the results of our theory for FJC, in
order to identify the major model-dependent features. The freely joint chain
consists of discreet bonds of length $a=2l_{p}$, whose orientations are
mutually independent. The corresponding loop free energy can be written as, 
\begin{equation}
\exp \left[ -F_{loop}(l)\right] \simeq 2\left( \frac{3}{2\pi }\right) ^{3/2}%
\frac{v_{0}}{l_{p}^{2}}\sum\limits_{n=1}^{\infty }\frac{\delta \left(
l-2nl_{p}\right) }{n^{3/2}}
\end{equation}%
The prefactor here ensures that the asymptotic behavior of the free energy
at the large--loops limit coincides with the WLC result, given by Eq. (\ref%
{Floop}). Now, one can find $W\left( q\right) $ for FJC model:%
\begin{equation}
W\left( q\right) =\frac{3\alpha }{2\sqrt{\pi }l_{p}^{2}}\sum\limits_{n=1}^{%
\infty }\frac{\exp \left( 2ql_{p}n\right) }{n^{3/2}}
\end{equation}

For negative $\epsilon $, the change of the elastic description results in a
modest shift of the critical tension $q^{\ast }$ at which the unfolding
transition occurs: 
\begin{equation}
q_{FJC}^{\ast }\simeq \frac{\epsilon }{2}+\frac{1}{l_{loop}(\epsilon /2)}%
\simeq \frac{\epsilon }{2}+\frac{1}{2l_{p}}
\end{equation}%
This should be compared to WLC result, Eq. (\ref{qstarwlc}).

\begin{figure}[tbp]
\begin{center}
\includegraphics[
height=4.5775in,
width=3.2681in 
]{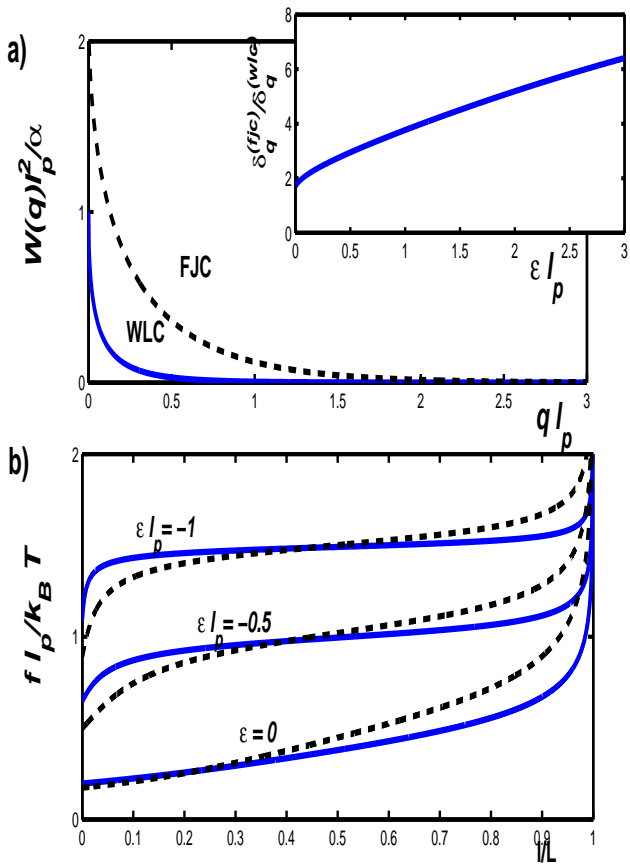}
\end{center}
\caption{a) Comparison of $W(q)$ functions calculated for WLC and FJC
models. Insert: relative width $\protect\delta _{q}$ of unzipping
crossovers, for FJC and WLC, as a function of basepairing energy. b)
Unzipping behavior of WLC (solid) and FJC (dashed) systems for the
basepairing energy changing between $\protect\epsilon =0$ $\ $\ and $\protect%
\epsilon =-k_{B}T/l_{p}$. For the both models, we put $\protect\alpha =0.01$%
. }
\label{wq}
\end{figure}

While the position of the unzipping point, $q_{el}=\epsilon /2$, is model
independent, the width of the crossover is very sensitive to the behavior of 
$W\left( q\right) $. As one can see in Figure \ref{wq}, the shapes of $%
W\left( q\right) $ curves are substantially different for WLC and FJC
models. This difference arrises because of the relative suppression of the
short loops in WLC case. What is especially remarkable is that the width of
the unzipping crossover, $\delta _{q}=\sqrt{2W\left( \epsilon /2\right) }$
decreases with $\epsilon $ in a strongly model dependent manner. In fact,
upon change of $\epsilon l_{p}$ from $0$ (dsDNA denaturation point) to $3kT$
(which \ roughly corresponds to physiological conditions for --GCGCGC--
sequence), the ratio of $\delta _{q}$ for the two models changes by factor
of $\simeq 3.5$. Therefore, an experimental study of the unzipping crossover
at variable conditions (e.g. temperature), would open a new possibility of
testing the plausible models of ssDNA elasticity.

\section{Conclusions}

In this paper, we have discussed the effects of basepairing on the
stretching behavior of ssDNA, within a theoretical framework compatible with
an arbitrary underlying model of chain elasticity. Our conclusion is that in
a generic case, the stretching curves exhibits two related but distinct
feature: the second--order \emph{unfolding} phase transition, and the sharp 
\emph{unzipping }crossover. The latter is reminiscent of the first--order
transition in dsDNA, as well as to the mechanical response of non-random RNA
molecules\cite{Hwa}. On the other hand, we have interpret the unfolding as a
topological transition. At the critical point, the typical size of
independently folded domains diverges (in the thermodynamic limit). This
transition is due to the competition of conformational entropy and elastic
free energy, and it is expected to occur even in the regime when basepairing
is energetically unfavorable.

In the light of our results, one can see a clear relationship between the
three types of the force-induced denaturation: (i) unzipping of dsDNA\cite%
{unzip1}-\cite{unzip}, (ii) denaturation of RNA with a preferred secondary
structure\cite{Hwa}, and (iii) stretching a self--complimentary or random
ssDNA. While in the case of dsDNA the unzipping occurs as a first order
transition, it becomes a crossover for the two other cases. The width of the
crossover is defined by the typical length of a single hybridized region. We
expect this width to be sensitive to the sequence of ssDNA/RNA. In fact, the
force-induced denaturation of RNA \cite{Hwa} was predicted to show a
sequence of unzipping steps. It follows from our discussion that the
sharpness of those steps may be even more pronounced than was originally
predicted within FJC-based model. In the case of ssDNA, the sequence
disorder must result in smearing of the unzipping crossover. Because of \
its entropic nature, the unfolding transition is expected only in the case
of ssDNA (or long RNA, in molten phase \cite{hwa-RNA}).

At present, the experimental indications of the second order unfolding
transition are not conclusive enough \cite{ssDNAexp}-\cite{ssDNAexp1}. On
the other hand, experiments with uniform self--complimentary DNA show a
clear manifestation of the sharp unzipping crossover \cite{ssDNAexp2}.
However, their precision is still insufficient to make a quantitative
comparison with the theory, and to distinguish between different underlying
elastic models. Based on our theory, one may extract this information by
performing a systematic experimental study of the unzipping behavior for
various values of hybridization energy (e.g. various temperatures). As we
have shown, the width of the crossover is very sensitive to the energy cost
of the hairpin loop.

\begin{acknowledgments}
\textbf{Acknowledgments.} The author thanks B. Shraiman, D. Lubensky,
J.Marko, and E. Siggia for valuable discussions.
\end{acknowledgments}

\end{document}